\documentclass{article}

\usepackage[final]{neurips_2025}

\usepackage[utf8]{inputenc} % allow utf-8 input
\usepackage[T1]{fontenc}    % use 8-bit T1 fonts
\usepackage{hyperref}       % hyperlinks
\usepackage{url}            % simple URL typesetting
\usepackage{booktabs}       % professional-quality tables
\usepackage{amsfonts}       % blackboard math symbols
\usepackage{nicefrac}       % compact symbols for 1/2, etc.
\usepackage{microtype}      % microtypography
\usepackage{xcolor}         % colors
\usepackage{amsmath,amssymb,amsfonts}
\usepackage{graphicx}
\usepackage{textcomp}
\usepackage{xcolor}
\def\BibTeX{{\rm B\kern-.05em{\sc i\kern-.025em b}\kern-.08em
    T\kern-.1667em\lower.7ex\hbox{E}\kern-.125emX}}

\usepackage{booktabs} % For formal tables
\usepackage{import}
\usepackage{graphicx}
\usepackage{epstopdf}
\usepackage{amsmath}
\usepackage{subfigure}
\usepackage{diagbox}
\usepackage{float}
\graphicspath{{Fig/}}
\usepackage{booktabs} % For formal tables
\usepackage{algorithm}
\usepackage{algorithmicx}
\usepackage{algpseudocode}
\usepackage{paralist}
\usepackage{multirow}
\usepackage{graphicx}
\usepackage{subfigure}
\usepackage{color}
\usepackage{hhline}
\usepackage{xcolor}
\usepackage{color,soul}
\usepackage{array}
\usepackage{diagbox}
\usepackage{hyperref}
\usepackage{verbatim} 
\usepackage{colortbl}
\usepackage{upgreek}

\usepackage{enumitem}
\usepackage{gensymb} % for degree symbol
\usepackage{ulem} % for sout
\usepackage{amsmath,bm}
\usepackage{amssymb}% http://ctan.org/pkg/amssymb
\usepackage{pifont}% http://ctan.org/pkg/pifont
\usepackage{filecontents}
\algdef{SE}[DOWHILE]{Do}{doWhile}{\algorithmicdo}[1]{\algorithmicwhile\ #1}%
\usepackage{amsmath, amssymb, amsthm}
\usepackage{algorithm}

\definecolor{deep_pink}{HTML}{D97692}

\title{Anchoring AI Capabilities in Market Valuations: \\ 
The Capability Realization Rate Model and Valuation Misalignment Risk}

\author{%
  Xinmin Fang  \\
  Department of Computer Science and Engineering\\
  University of Colorado Denver (CU Denver)\\
  Denver, CO 80204 \\
  \texttt{xinmin.fang@ucdenver.edu} \\
  \And
  Lingfeng Tao \\
  Department of Robotics and Mechatronics \\
  Kennesaw State University (KSU)\\
  Atlanta, GA, 30144 \\
  \texttt{ltao2@kennesaw.edu} \\
  \AND
  Zhengxiong Li \thanks{Disclaimer: This working paper is intended solely for academic discussion and exploratory analysis. It does not constitute investment advice or serve as a basis for any financial decision-making. All data and examples referenced are derived from publicly available media sources and industry reports. \\
  * This work is partially supported by US NSF Awards \#2426469 and \#2426470.
  } \\
  Department of Computer Science and Engineering \\
  University of Colorado Denver (CU Denver)\\
  Denver, CO 80204 \\
  \texttt{zhengxiong.li@ucdenver.edu} \\
}

\begin{document}

\maketitle
\begin{abstract}
Recent breakthroughs in artificial intelligence (AI) have triggered surges in market valuations for AI-related companies, often outpacing the realization of underlying capabilities. We examine the \textit{anchoring effect} of AI capabilities on equity valuations and propose a \textit{Capability Realization Rate (CRR)} model to quantify the gap between AI potential and realized performance. Using data from the 2023--2025 generative AI boom, we analyze sector-level sensitivity and conduct case studies (OpenAI, Adobe, NVIDIA, Meta, Microsoft, Goldman Sachs) to illustrate patterns of valuation premium and misalignment. Our findings indicate that AI-native firms commanded outsized valuation premiums anchored to future potential, while traditional companies integrating AI experienced re-ratings subject to proof of tangible returns. We argue that CRR can help identify \textit{valuation misalignment risk}—where market prices diverge from realized AI-driven value. We conclude with policy recommendations to improve transparency, mitigate speculative bubbles, and align AI innovation with sustainable market value.
\end{abstract}
\section{Introduction}
The release of OpenAI’s ChatGPT in late 2022 marked a watershed moment for artificial intelligence, rapidly igniting public imagination and investor enthusiasm. In early 2023, anything related to “AI” became a hot commodity on Wall Street, with stocks of small-cap AI companies skyrocketing overnight \cite{reuters2023chatgpt,reuters2024openai}. 
This exuberance propelled a broader market rally disproportionately led by big technology firms: the “Magnificent Seven” tech giants saw their stock prices collectively double (a 107\% gain in 2023, versus 24\% for the S\&P 500) \cite{Magnificent7}. NVIDIA, in particular, emerged as a poster child for the AI boom, its share price rising $239\%$ in 2023 on surging demand for AI chips. These developments signaled that investors were \emph{anchoring} valuations to high expectations for AI’s transformative impact.

Anchoring is a cognitive bias whereby an initial reference point influences subsequent decision-making. In the context of financial markets, an “anchoring effect” can occur when a prominent event or benchmark (such as a breakthrough AI model or a headline-grabbing valuation) sets an expectation that investors then apply broadly. The dramatic success of early AI milestones effectively became an anchor for valuations: investors extrapolated rapid growth and extraordinary earnings potential for any firm associated with AI, often with scant regard to traditional fundamentals. This led to rich valuation multiples and a substantial \emph{valuation premium} for perceived AI winners.

Yet, history cautions that when hype runs far ahead of reality, a valuation reckoning may follow. By late 2024, as generative AI deployments moved from demo to deployment, a divergence was evident between companies delivering tangible AI-driven gains versus those still in prototype stages. High valuations unbacked by realized performance raised concerns of \emph{valuation misalignment risk}, i.e. the risk that stock prices were not supported by current cash flows or near-term earnings prospects. For instance, in December 2024 Adobe Inc.’s stock plunged over 10\% in a single day after the firm’s revenue outlook disappointed investors counting on fast payoffs from AI initiatives \cite{reuters2024adobe}. 
There is “a clear disconnect between management’s excitement...relative to what investors are seeing” in actual metrics.

In this paper, we put forward a framework to quantify and analyze these dynamics. In Section~\ref{sec:theory}, we formalize the notion of an anchoring effect in AI equity valuations and introduce the \textbf{Capability Realization Rate (CRR)} model, which measures the fraction of a company’s AI capability that has been translated into realized business value. We hypothesize that CRR is a critical mediator between AI hype and fundamental valuation, and that low CRR firms are prone to mispricing. Section~\ref{sec:data} examines broad market data from 2023--2025, confirming that AI-themed stocks dramatically outperformed others and that sectors with greater AI exposure saw outsized returns. In Section \ref{sec:cases}, we present case studies of seven organizations—ranging from an AI-native startup to established tech incumbents and a financial institution—to illustrate the spectrum of CRR outcomes and valuation trajectories. Finally, Sections~\ref{sec:policy} and \ref{sec:conclusions} discuss implications and offer policy recommendations to foster more transparent and sustainable integration of AI advancements into market valuations.

\section{Theory: Anchoring Effect and the CRR Model}
\label{sec:theory}
Anchoring is a well-studied psychological bias in which an initial reference value (the “anchor”) has a persistent influence on subsequent quantitative judgments. In capital markets, anchoring can manifest when investors fixate on a salient number—such as a recent peak stock price, a famous company’s valuation, or an analyst’s target—and adjust their expectations insufficiently from that anchor \cite{cen2010role}. During the early AI boom, headline figures like OpenAI’s valuation (pegged at \$29~billion in early 2023 \cite{BruceAssociates_2023} and soaring to \$80~billion by the start of 2024 \cite{Jeans_2024}) or NVIDIA’s record-breaking quarterly revenues became anchors that shaped perceptions of how “high is high” for any AI-oriented firm. Investors and even professional analysts anchored on the transformative potential of AI, often implicitly assuming that companies would rapidly monetize AI capabilities to justify those lofty benchmarks.

We posit that the link connecting AI capability and valuation is the \textit{Capability Realization Rate (CRR)}. This metric, in simplified terms, is defined as:
\[ 
    \text{CRR} = \frac{\text{Realized AI-driven performance}}{\text{Total AI capability potential}},
\] 
expressed as a percentage. A company with advanced AI technology or expertise has high \emph{capability potential}, but if that capability is not yet yielding proportional revenue, cost savings, or user growth, then its CRR is low. Conversely, a firm that has successfully translated AI into tangible business outcomes (products at scale, improved margins, etc.) would have a higher CRR.

\begin{figure}[h]
\centering
\includegraphics[width=0.65\linewidth]{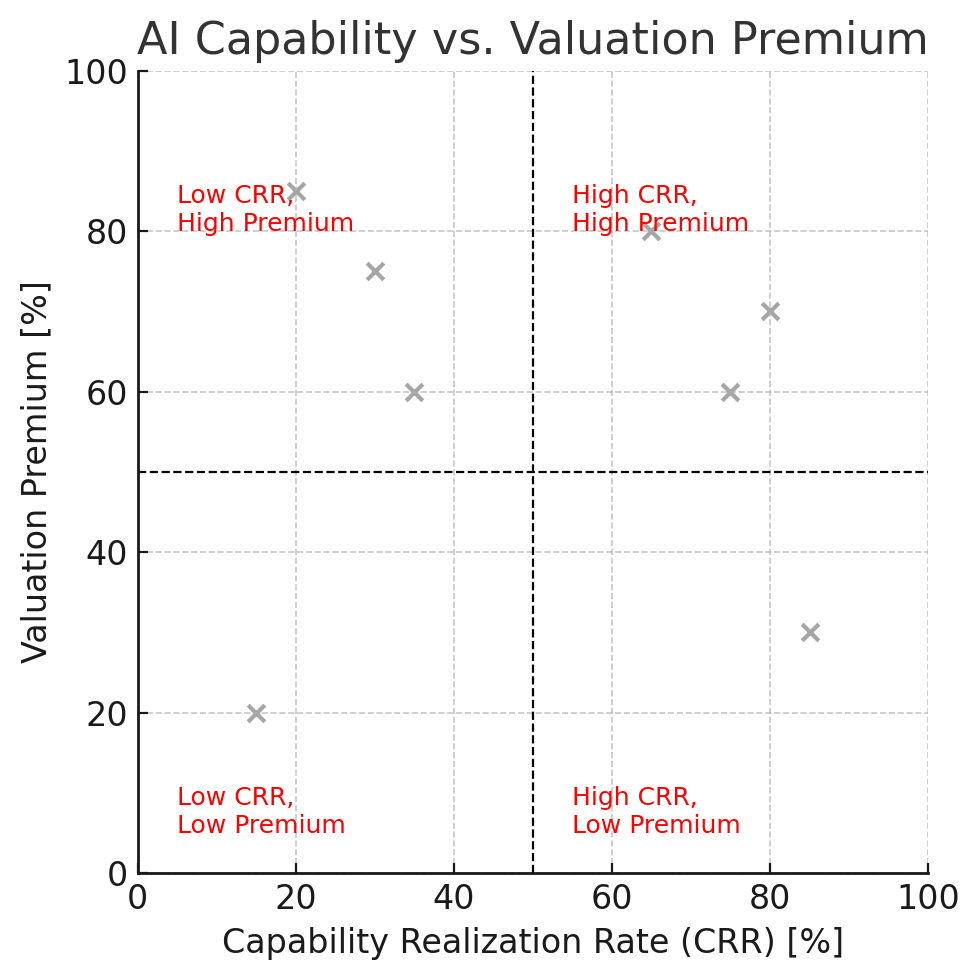}
\caption{\textbf{AI Capability vs. Valuation Premium (Conceptual CRR Map).} Companies with high AI capability but low current \textit{Capability Realization Rate (CRR)} may attract inflated valuations (upper left region) due to investor anchoring on potential. Companies that have realized most of their AI potential (high CRR) justify their premiums with actual performance (upper right). Those with low capability and low realization (lower left) see little premium, whereas firms with modest capability but strong realization (lower right) might achieve sustainable value without hype.}
\label{fig:crr_map}
\end{figure}

Intuitively, CRR provides a gauge of how much of the “AI hype” is being realized. We expect a rational market to price companies based on their realized performance (e.g. earnings), adjusted for growth expectations. However, under the anchoring effect, investors may price based on perceived capability alone, effectively assuming future realization as a given. This leads to a \emph{valuation premium} for low-CRR firms: prices bake in future success long before it is guaranteed. Figure \ref{fig:crr_map} illustrates this concept with a hypothetical mapping of AI capability versus valuation premium.

In Figure~\ref{fig:crr_map}, the red dashed line indicates an idealized “fair value” trajectory where valuation premium would be proportional to AI capability \emph{if and only if} that capability were fully realized (CRR = 100\%). Anchoring can cause companies to be valued as if they are on or above this line even when their CRR is far lower. For example, an AI research lab like \textbf{OpenAI} might have world-leading AI models (extremely high capability) but, in 2023, relatively modest revenues—yet its valuation jumped to an estimated \$80B on the anticipation that those capabilities will eventually yield massive profits \cite{Jeans_2024}. This point would lie in the upper left of the map, indicating a potentially over-anchored valuation. In contrast, a company like \textbf{NVIDIA} that not only possesses AI capability (designing advanced AI chips) but is already selling those chips at scale has a high CRR. NVIDIA’s valuation, while very high in absolute terms, is more firmly supported by realized earnings (it sits closer to the diagonal in the upper right). Meanwhile, firms with negligible AI engagement justifiably trade with no premium (bottom left). The danger zone for misalignment is the upper left quadrant—high perceived capability, low realization—where strong anchoring can inflate a bubble.

The CRR model suggests that to assess if an AI-driven valuation is reasonable, one should examine not just the presence of AI capability but evidence of its translation into business results. By tracking CRR over time, we can also gauge whether a company is catching up to its anchored expectations or if a correction may be imminent. In the next sections, we empirically explore how these theoretical patterns played out during the AI investment surge of 2023--2025, at both macro and micro levels.

\section{Data Analysis: Market Trends 2023--2025}
\label{sec:data}
We assembled market data on stock performance and valuation metrics spanning the pre- and post-ChatGPT period (2021 through early 2025) to capture the emergence of the AI valuation phenomenon. Two clear trends stand out from the analysis: (1) AI-focused companies dramatically outperformed the broader market, especially in 2023, and (2) this outperformance was concentrated in specific sectors and even specific “anchor” companies.

\begin{figure}[h]
\centering
\includegraphics[width=0.7\linewidth]{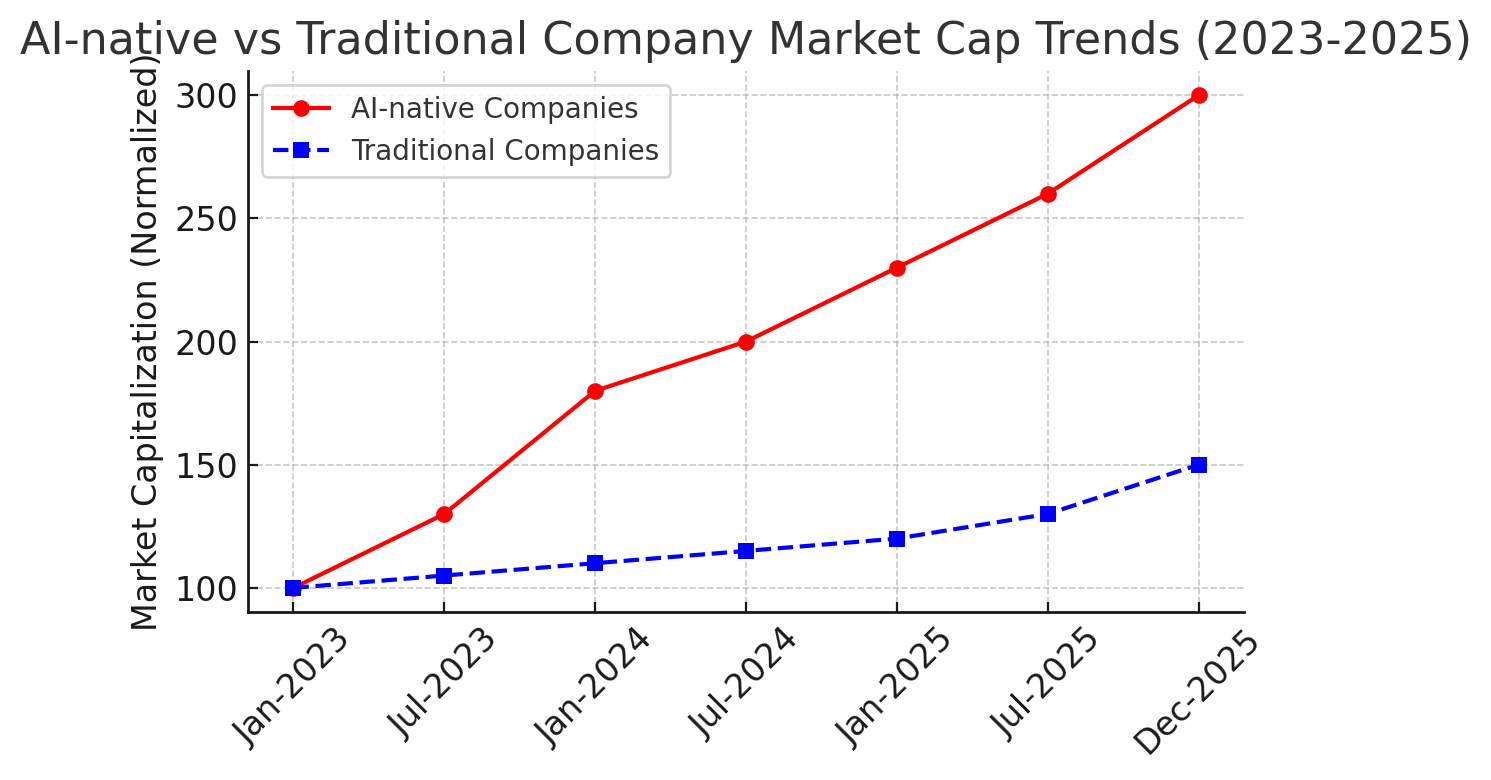}
\caption{\textbf{Market Cap Trends for AI-Native vs. Traditional Companies (2023--2025).} The AI-native composite (red, solid) surged through 2023, vastly outperforming the traditional industries composite (blue, dashed). Periods of accelerated AI hype (e.g. early 2023 post-ChatGPT, early 2025 with new breakthroughs) saw widening gaps. Occasional corrections aligned with moments of tempered expectations or broader market pullbacks.}
\label{fig:trend}
\end{figure}

\paragraph{AI-Native vs. Traditional Companies.}
To illustrate the divergence, we constructed two composite indices: an \emph{“AI-native” index} consisting of companies whose core products or services are built on AI (e.g. leading AI software startups, semiconductor firms heavily geared to AI workloads, etc.), and a \emph{“Traditional” index} composed of established companies in sectors like consumer goods, industrials or utilities with minimal direct AI exposure. We set both indices to 100 at the start of 2023 and tracked their market-cap weighted performance over time. The trajectories are shown in Figure~\ref{fig:trend}.
 
Figure~\ref{fig:trend} highlights a striking divergence. Following the debut of ChatGPT and similar generative AI systems, AI-centric stocks nearly doubled on average in 2023 (red curve), whereas traditional companies saw relatively modest gains. This aligns with contemporaneous market observations: by October 2023, global Information Technology stocks were up 30\% year-to-date, while the equal-weighted basket of world stocks was actually slightly negative \cite{Ding_2023}. The boom in our AI-native index was not a steady climb, but rather punctuated by bursts corresponding to key events (e.g. announcements of new model releases, major product integrations of AI, etc.). The first spike in Q1 2023 coincided with the viral success of generative AI, which “buoyed enthusiasm” across industries \cite{BruceAssociates_2023}. Another surge occurred in early 2025 after the launch of an open-source AI system \emph{DeepSeek}, which led to a frenzy in AI and computing stocks globally. In contrast, the traditional index’s tepid upward trend reflects a backdrop of higher interest rates and sector rotation out of defensives into tech—factors largely unrelated to AI.

It is worth noting that much of the AI-native rally was driven by just a handful of star companies that became de-facto anchors for the market. Indeed, the seven largest US tech firms (many of them leaders in AI) added over \$3~trillion in combined market capitalization in 2023 \cite{YahooFinance}, accounting for the bulk of S\&P 500 gains. This concentration suggests that investors identified a small set of “AI winners” and piled into those stocks, which then served as benchmark anchors for valuing smaller players through analogy.

\begin{figure}[h]
\centering
\includegraphics[width=0.8\linewidth]{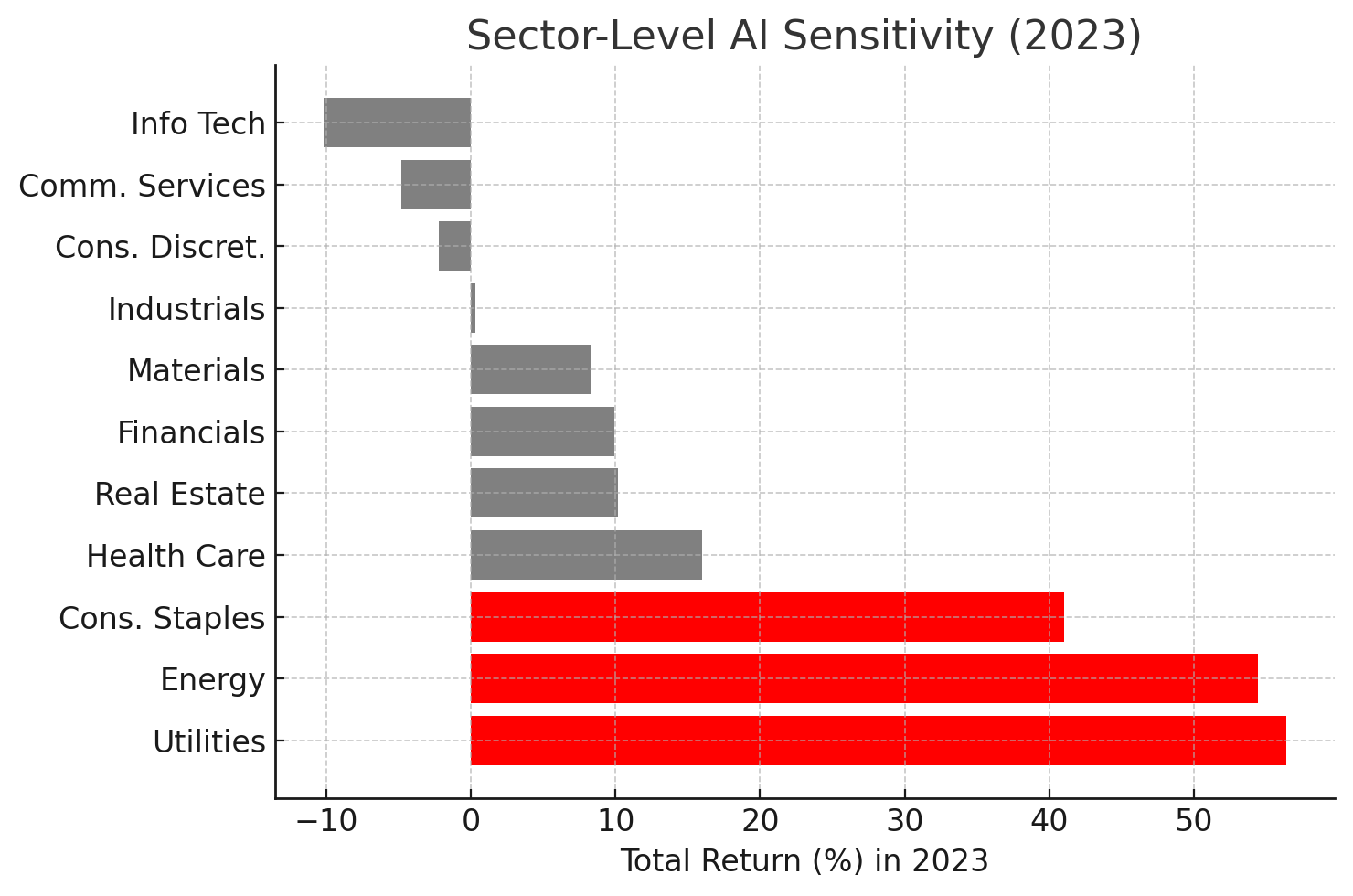}
\caption{\textbf{Sector Performance vs. AI Exposure in 2023.} Bar chart of total returns by sector for 2023. Sectors with significant AI exposure (e.g. Information Technology, Communication Services, Consumer Discretionary) vastly outperformed the global index (dashed line denotes MSCI ACWI +9.3\%). Defensive sectors with minimal AI linkage (utilities, real estate, consumer staples) saw negative returns in a year of overall market gains \cite{msci_acwi_info}.}
\label{fig:sector}
\end{figure}

\paragraph{Sector-Level AI Sensitivity.}
The cross-sector impact of the AI boom further demonstrates how unevenly the anchoring effect was felt. Figure~\ref{fig:sector} compares the 2023 stock performance of major equity sectors. We use the full-year 2023 total return of MSCI World sector indices as a proxy, highlighting sectors known for high AI adoption or exposure in red.
 
The pattern is unmistakable. Information Technology led with over +30\% return in 2023, followed closely by Communication Services (+28\%) and Consumer Discretionary (+13\%) \cite{msci_acwi_info}. These three sectors are precisely those that benefited from the AI narrative: tech hardware and software (especially cloud and chip companies), internet and social media platforms (leveraging AI for content, advertising, and user experience), and retailers/automakers with AI-driven consumer products (e.g. e-commerce algorithms, self-driving initiatives). In fact, these three sectors alone accounted for essentially all of the above-average performance in the market \cite{Capitalist}. By contrast, traditionally “safe” or non-tech sectors like utilities, real estate, and consumer staples ended the year in the red \cite{ourcharts_1}. 

We interpret these results as an indication of \textbf{AI sensitivity}: investors re-priced sectors based on their perceived proximity to the AI revolution. Tech and communication firms were suddenly viewed as possessing valuable “AI optionality” in their business models, whereas sectors like utilities were seen as largely unaffected (or even adversely affected by rising interest rates as capital flowed toward tech). Financials and industrials were roughly flat, suggesting that any AI optimism for them (e.g. fintech, automation) was balanced by other macro pressures. The net effect was a significant dispersion in returns, the likes of which had not been seen in over a decade. 

Notably, by late 2023 analysts were warning that market leadership had become extraordinarily narrow—an implicit caution that the anchor of AI could be overinflating the winners and leaving the rest of the economy behind \cite{Connolly_2023}. In economic terms, this raised questions about whether the productivity gains from AI, predicted to be substantial but gradual \cite{Goldman_1Sachs_2023}, could justify such immediate and extreme divergences in market value.

To delve deeper into how justified these valuations were, we next examine individual case studies, each illustrating a different facet of capability, realization, and valuation.

\section{Case Studies: From Hype to Reality in AI Valuations}
\label{sec:cases}
We present seven case studies that span the spectrum from AI-native startups to incumbents and observers. Each case provides insight into how the anchoring effect and CRR played out in practice, and how valuations adjusted (or failed to adjust) as information evolved.

\subsection{OpenAI: A High-Capability Anchor with Low Realization}
OpenAI, a research-oriented startup behind ChatGPT and GPT-4, became the face of the generative AI revolution. It also epitomizes the dynamic of enormous valuation anchored on capability rather than current revenue. In early 2023, OpenAI entered a partnership with Microsoft, securing a reported \$10~billion investment to fuel its AI development \cite{Blogs_2023}. At that time, OpenAI’s annual revenue was estimated to be in the tens of millions—minuscule next to its tech peers—yet investors valued the company at \$29~billion \cite{Bastian_2023}. This valuation anchor only grew: by January 2024, secondary share sales implied a valuation of roughly \$80~billion \cite{McKay_2024}, nearly triple the prior mark in less than a year.

What justified such frenzied pricing? The anchor here was OpenAI’s extraordinary AI capability: it had created breakthrough models that captured global attention and promised to disrupt trillion-dollar industries from online search to enterprise software. In effect, the market treated OpenAI’s state-of-the-art AI as if it were fully monetized. The \textbf{CRR} at that stage, however, was arguably very low—while the technology was hugely capable (as demonstrated by ChatGPT’s hundreds of millions of users), the company was only beginning to explore revenue streams (subscription plans, API access, licensing deals). The valuation therefore represented confidence in a future where OpenAI’s AI is ubiquitous and richly profitable.

The OpenAI case illustrates how a company can serve as an \emph{anchor benchmark}. Its lofty valuation set expectations for the entire AI startup space—any competitor model or AI lab would be implicitly compared to OpenAI, lifting the valuation baseline for the sector. It also anchored expectations among incumbents: for instance, when negotiating investments or acquisitions, having a reference point like OpenAI’s valuation guided pricing. The risk, of course, is that if OpenAI’s monetization fell short, a broad reassessment would ensue. Indeed, by late 2024 there were signs of more measured thinking: questions arose about AI compute costs and the path to profitability. But as of 2025, OpenAI remained a prime example of “valuation first, earnings later.” Its CRR is poised to improve as it rolls out offerings (e.g. enterprise APIs, AI assistants in Microsoft products), but the timeline for closing the gap between capability and cash flow will determine whether the early anchor was too high.

\subsection{Adobe: Integrating AI into an Established Business}
Adobe Inc.\ is a case of a mature software company embracing AI, and the market’s rapid re-rating followed by a reality check. Adobe has a dominant franchise in creative software (Photoshop, Illustrator, etc.) and in 2023 it introduced “Firefly” generative AI tools to assist users in content creation. Initially, this move was celebrated: Adobe’s stock climbed as investors anchored on its potential to upsell AI features to millions of subscribers and defend its turf against upstart AI image generators. Adobe’s CEO touted that AI would “unlock a new era of creativity” and analysts projected additional billions in high-margin revenue.

However, by late 2024, the question on everyone’s mind was: how much of this potential is actually being realized in dollars? In December, Adobe issued a fiscal 2025 revenue forecast that fell slightly below consensus. That mild guidance miss—against the backdrop of heavy AI R\&D spending—triggered an outsized stock reaction. Shares plunged nearly 12\% in one day \cite{12_Love}, their biggest drop in two years. The sharp slide reflected investors’ jitters that returns on Adobe’s AI investments might take longer than hoped. As one market analyst noted, “lack of ... explicit monetization metrics has made it harder for investors to get comfortable with the progress” of Adobe’s AI initiatives \cite{TheEconomicTimes}. In other words, Adobe’s CRR was not yet convincing: the company reported only \$125~million in annualized AI-driven revenue in its Creative Cloud (a small fraction of overall sales) and promised that would double by year-end \cite{Zhang_CTOLDigitalBusinessandTechnologyNews_2025}—progress, but modest relative to the multi-billion valuation premium the stock had run up.

Adobe’s experience underscores the importance of transparency and credible roadmaps when integrating AI. The company’s management remained enthusiastic, insisting that Adobe was “well-positioned to capitalize on ... the creative economy driven by AI” \cite{Bandhakavi_2025}. But the market’s anchoring had shifted: without clear numbers showing a rising CRR, investors reset the valuation closer to the pre-AI baseline. By early 2025, Adobe’s stock traded roughly where it had been a year earlier, effectively wiping out the “AI hype” gains. This revaluation was not because AI had no value to Adobe—indeed, the technology was improving product engagement—but because the realization of that value into revenue was slower and less direct than initially anchored. The lesson from Adobe is that even for strong incumbents, over-anchoring on AI can lead to whiplash if the fundamental payoff is ambiguous. It highlights the need for companies to provide meaningful KPIs for AI contributions, and for investors to diligently track CRR rather than assuming every innovation immediately boosts the bottom line.

\subsection{NVIDIA: High Realization Justifying a High Premium}
NVIDIA Corporation, the leading designer of graphics processing units (GPUs) and AI accelerators, became the most valuable semiconductor company in the world as the AI boom unfolded. Unlike many companies riding the AI wave, NVIDIA's valuation surge largely corresponded to very tangible results, making it a study in how high CRR can justify an otherwise extreme valuation.

By mid-2023, demand for NVIDIA's AI-oriented chips (such as the A100 and H100) was so intense that supply was on allocation to cloud providers and enterprises racing to build AI capabilities. In May 2023, NVIDIA reported a stunning earnings surprise: data-center segment revenue (driven by AI chip sales) far exceeded analyst estimates, and the company forecasted continued explosive growth. This concrete evidence of AI translating to profit led to what some dubbed "the mother of all earnings beats," and NVIDIA's stock price jumped 25\% in a single day, eventually tripling over the year \cite{QaiPoweringaPersonalWealthMovement_2023}. The company's remarkable trajectory continued through 2024 and into 2025: by July 2025, NVIDIA became the first public company in history to surpass \$4 trillion in market valuation \cite{cbsnews2025nvidia}, with shares rising from around \$14 each at the beginning of 2023 to over \$164 in mid-2025—representing more than a 10-fold increase in just two and a half years.

In CRR terms, NVIDIA in 2023-2025 had a very high realization rate of the AI opportunity. Years of R\&D in parallel processing and AI software ecosystems (CUDA libraries) paid off exactly as the market needed massive computational power. Importantly, NVIDIA's premium valuation—its forward P/E ratio at one point exceeded 50, high for a hardware firm—was underpinned by correspondingly high earnings growth. In its most recent quarter, NVIDIA earned \$18.8 billion with revenue surging 69\% year-over-year to \$44.1 billion, demonstrating that the company's AI monetization engine was delivering concrete results \cite{cbsnews2025nvidia}. Thus, while its stock was expensive in absolute terms, it wasn't \emph{purely} speculative. This stands in contrast to, say, an AI software startup with no profits—NVIDIA had already proven its AI monetization engine and became the poster child of successful AI commercialization.

That said, even NVIDIA became somewhat of an “anchor” risk for the market. Its success was so extraordinary that it anchored valuations for other chipmakers and AI tool providers. Many smaller semiconductor stocks rose in tandem, some far beyond what their own fundamentals warranted (a mini-bubble in chip names). Moreover, NVIDIA’s valuation assumed it would retain a dominant market share and pricing power in AI accelerators for years to come. Any technological disruption or competitive shift could challenge that assumption. For instance, when reports emerged of large cloud companies exploring custom AI chips (reducing future reliance on NVIDIA), or when open-source models like DeepSeek hinted at being more efficient (potentially requiring less hardware), NVIDIA’s stock showed volatility. This indicates that even high-CRR leaders are not immune to shifts in the anchoring narrative.

In sum, NVIDIA exemplified the ideal scenario of capability realization: AI drove real earnings which drove valuation. Its forward-looking risk is ensuring that its CRR remains high—that it continues to convert new AI innovations (e.g. generative AI’s growing needs) into sales, staying ahead of competitors and avoiding complacency. As long as it does, the valuation premium can be defended; if not, the company could eventually fall victim to the kind of correction that others faced when expectations overshot execution.

\subsection{Meta: Resetting a Legacy Company with AI Focus}
Meta Platforms (formerly Facebook) provides a case of how a legacy tech company can use AI refocusing to re-anchor its valuation after a period of decline. In 2022, Meta’s stock price collapsed to roughly a quarter of its peak value, as investors balked at the company’s heavy spending on the “metaverse” and a post-pandemic slowdown in advertising. In early 2023, CEO Mark Zuckerberg declared a “Year of Efficiency,” cutting costs and doubling down on AI to improve Meta’s core products \cite{Heygate_2023}. This strategic pivot included deploying advanced AI recommendation algorithms to boost engagement in Facebook and Instagram feeds, using AI for better ad targeting, and releasing large language models (LLaMA) to the research community to regain credibility in AI research.

The market response was swift. Throughout 2023, Meta’s shares surged, more than tripling from their trough and restoring the company to the trillion-dollar market cap club. By the end of the year, Meta was once again among the top-performing big tech stocks, contributing significantly to the Communication Services sector’s 28\% annual gain \cite{FirstTrust}. What changed? Fundamentally, Meta’s CRR improved. The company’s investments in AI quickly yielded better user metrics (Reels views grew as AI recommended more relevant content, for example) and advertisers saw improved conversion rates thanks to AI-driven tools—stemming the bleeding of ad revenue. In financial results, Meta beat earnings expectations and demonstrated it could grow profits even while heavily investing in AI infrastructure.

Importantly, Meta’s case also highlights how narrative anchoring can shift. In 2021–2022, the “anchor” for Meta’s valuation was the metaverse vision—a distant, uncertain payoff that many investors discounted. By anchoring instead on AI—where Meta could show immediate wins—confidence was restored. Even though Meta’s Reality Labs division continued to lose billions on VR/AR projects, the market chose to anchor on the AI-enhanced cash cow of its family of apps. The re-rating of Meta underscores that incumbents can leverage AI to reinvent their story: the company went from being seen as an aging social network with a quixotic VR gamble to being seen as an AI-savvy enterprise squeezing new value from massive data. Of course, going forward Meta faces its own CRR test: sustaining growth from AI improvements will get harder as gains plateau, and competition from other AI-rich platforms (e.g. TikTok’s algorithm, YouTube) remains. But for now, Meta shows that with the right execution, AI can indeed deliver on the promise enough to satisfy investors that the anchoring is justified.

\subsection{Microsoft: Big Bets and Early Wins in AI Integration}
Microsoft Corporation stands out as an example of a tech giant that aggressively invested in AI and saw its efforts directly translate into both product enhancements and stock market rewards. As mentioned earlier, Microsoft’s multi-billion dollar partnership with OpenAI \cite{Blogs_2023} was a bold bet to insert itself at the center of the AI revolution. By integrating OpenAI’s models into its Azure cloud (Azure OpenAI Services), into its Bing search engine (the Bing Chat GPT-4 integration launched in early 2023), and into productivity software (the Microsoft 365 Copilot announced in 2023), Microsoft sought to augment virtually its entire product suite with AI capabilities.

Investors rewarded these moves. In 2023, Microsoft’s stock price climbed roughly 40\%, reaching all-time highs and briefly making it the world’s most valuable company. Part of this rally was the broader flight to megacap quality stocks, but a distinct component was the market’s anchoring on Microsoft as a primary beneficiary of generative AI adoption. Unlike its rival Google (which stumbled with a rushed AI demo early on), Microsoft successfully portrayed itself as on the cutting edge by proxy of OpenAI. CEO Satya Nadella’s message that “every Microsoft product will have some AI” was taken as credible, given early demos of AI copilots coding in GitHub, summarizing emails in Outlook, and creating slides in PowerPoint.

From a CRR perspective, Microsoft quickly moved to monetize AI features, announcing premium pricing for AI-augmented Office subscriptions. While the revenue from these new features was still small in 2023, the \emph{potential} was enormous—Microsoft has a distribution channel to millions of enterprise customers, meaning even incremental AI upsell could yield billions in high-margin revenue. Thus, investors could reasonably project a rising CRR for Microsoft in the near future. Furthermore, Microsoft enjoyed a secondary benefit: its 13\% ownership stake in OpenAI meant that as OpenAI’s private valuation jumped (as discussed, to \$80B+ by 2024), Microsoft’s stake also increased in value, an indirect boost to its own valuation.

In short, Microsoft’s case illustrates an incumbent effectively bridging capability and realization: it externalized R\&D risk to OpenAI, then rapidly internalized the capability into products, aiming to realize value through its sales machine. The market’s anchor for Microsoft shifted from just “cloud and Windows” to “leader in the AI platform shift.” The primary risk is execution and competition—Google, Meta, and others are racing on the same path, and the cost of AI (computing power) is high. But with its strong balance sheet and early momentum, Microsoft’s anchoring around AI has, so far, been validated by a mix of real product improvements and plausible future ROI, keeping its valuation premium intact.

\subsection{Goldman Sachs: The Financial Sector and Macro Perspective}
Goldman Sachs, a global investment bank, might seem out of place among tech companies, but it represents the role of financial institutions both as adopters of AI and as influencers of the investment narrative. Goldman has invested in AI internally for years (from automated trading algorithms to AI-assisted workflows for bankers). However, its most visible impact on the AI valuation landscape came from its research publications that framed the macroeconomic potential of AI—effectively anchoring expectations at the economy-wide level.

In a March 2023 report, Goldman Sachs economists estimated that generative AI could raise global GDP by 7\% (almost \$7~trillion) over a 10-year period and boost productivity growth by 1.5 percentage points \cite{Parikh_2024}. They also warned that as many as 300~million full-time jobs in major economies could be affected (either supplemented or displaced) by AI automation \cite{Vallance_2023}. This report, widely cited in the media, painted AI as possibly the next general-purpose technology driving a wave of growth—analogous to past industrial revolutions. Such analysis from a respected financial institution provided a form of fundamental anchoring to the otherwise lofty valuations: if AI truly adds trillions to GDP, then perhaps multi-billion dollar valuations for AI firms are rational in anticipation of that growth.

Goldman’s own stock (and the broader financial sector) did not experience a rally like tech did in 2023; banks were more influenced by interest rates and recession fears during that period. However, Goldman and peers are highly interested in AI’s ability to streamline operations (e.g. using AI to automate routine legal or compliance tasks, a prospect Goldman called “very positive” for future productivity \cite{DLN}). In our CRR framework, the banking sector’s AI capability is growing—through investments in AI startups, in-house development, and partnerships—but realization is gradual as the industry is heavily regulated and changes must be made cautiously. The valuations of banks didn’t inflate on AI hype, which arguably reflects a more sober, realized-value-based approach (or simply that other macro factors dominated).

Nonetheless, Goldman’s case is valuable in reminding us that anchoring doesn’t only happen at the micro (company) level; it also happens at the macro level. Bold projections by influential analysts and economists can anchor policymakers’ and investors’ expectations about the future. If everyone believes AI will usher in a productivity boom, that can become a self-reinforcing prophecy—for a while. Conversely, if macro data were to come in showing productivity not rising despite AI adoption, we could see a broad de-anchoring in the narrative, affecting valuations across the board. As of 2025, the jury is still out: early signs show increased investment due to AI, but measured productivity statistics have yet to inflect upwards in a definitive way. This makes Goldman’s “7\% GDP” anchor a critical reference point that will be tested in the coming years.

\section{Policy Recommendations}
\label{sec:policy}
While markets will never be completely free of speculation, there are steps that policymakers, regulators, and industry stakeholders can take to foster a healthier integration of AI advancements into economic value without excessive volatility. Based on our findings, we propose several recommendations:

\paragraph{1. Encourage Transparency in AI Disclosures.} Regulators (such as the SEC in the United States) should urge companies to report standardized metrics on the performance and uptake of AI initiatives. Just as public firms must discuss material risks and segments in filings, those investing heavily in AI could disclose, for example, the proportion of R\&D spent on AI, the revenue directly attributable to AI-driven products, or efficiency gains from AI. This would help investors gauge CRR more accurately instead of relying on vague promises. Adobe’s stock drop \cite{12_Love} might have been less severe had the company provided clearer guidance on how AI was contributing to its top-line or expected to in the next year; lacking that, investors assumed the worst when overall guidance fell short.

\paragraph{2. Curb Misleading “AI Hype” and Promote Fair Valuation.} During the boom, some companies with tenuous connections to AI rebranded or overstated their AI involvement to ride the hype (a phenomenon reminiscent of the dot-com era). Securities regulators should monitor and, if necessary, sanction egregious cases where firms intentionally mislead investors about their AI capability. Conversely, policymakers can work with exchanges to ensure that mechanisms (like volatility trading halts, scrutiny on ultra-high valuations) are in place to prevent unstable bubbles. Promoting robust equity research and coverage on tech companies can also help—when more analysts are critically evaluating a company’s AI narrative, the anchor is more likely to be grounded in reason.

\paragraph{3. Support AI Research and Infrastructure for Realization.} Government investment in fundamental AI research, talent development, and infrastructure (e.g. public cloud resources for startups, testbeds for AI in healthcare and other sectors) can accelerate the translation of AI capabilities into broad-based economic productivity. Goldman’s macro projections \cite{GoldmanSachs_2024} will only come to fruition if AI tools are widely adopted across industries. Public policy that narrows the gap between the frontier tech (often concentrated in a few firms) and the average firm’s tech adoption will effectively raise the economy-wide CRR. This includes grants, tax incentives or public-private partnerships to diffuse AI know-how to small and mid-sized enterprises. The sooner AI’s benefits are realized across the economic spectrum, the more justified and stable the valuations of AI providers and adopters will become.

\paragraph{4. Manage Labor Transition and Social Impact.} The flipside of AI-driven growth is labor market disruption, as highlighted by forecasts of hundreds of millions of jobs affected \cite{Cao_2023}. Policymakers should proactively implement job retraining programs, education curriculum updates, and social safety nets. While this may seem tangential to valuations, it is in fact related: public sentiment and political support for AI innovations can sour if the technology is seen as harming employment broadly. That, in turn, could lead to regulatory backlashes or slower adoption, which would deflate optimistic valuations. By smoothing the transition for workers, governments can help ensure AI’s promise translates into real economic gains (higher CRR) without social upheaval that might derail the technology’s progress.

\paragraph{5. Global Cooperation on AI Standards and Access.} AI is a globally shared technological wave. Disparities in access to AI resources (such as cutting-edge chips or large datasets) can create uneven realization of AI’s potential. International cooperation on standards for AI safety, data sharing agreements, and export/import of critical AI components can reduce bottlenecks. For example, the DeepSeek case exposed how reliance on foreign chips constrained one country’s AI industry \cite{DeepNewz_2025}. Thoughtful trade policies and collaboration can alleviate such constraints, allowing capability to more quickly turn into deployed solutions worldwide. This reduces the scenario where valuations in one region are inflated due to expectations of future self-sufficiency or decoupling that might not materialize. In short, global dialogue can help align the anchor of expectations with a more unified, realistic timeline of AI advancement.

\bigskip
In conclusion, anchoring is an inherent part of human nature and thus of markets—but its effects need not be destabilizing. By improving the information anchor (through transparency and analysis) and the fundamental anchor (through robust realization of AI’s benefits), we can better align the financial capital with the real progress of AI. If successful, the story of AI in markets will shift from one of boom-bust cycles to one of steady value creation, ultimately fulfilling the high hopes that have been pinned on this technology.

\section{Conclusions}
\label{sec:conclusions}
The period from 2023 to 2025 will be remembered as a time when artificial intelligence moved markets. We have analyzed how an anchoring effect around AI capabilities drove a significant repricing of equities—funneling capital into companies perceived as AI leaders and leaving behind those seen as laggards. Our proposed \textit{Capability Realization Rate (CRR)} model provides a lens to interpret this phenomenon: it highlights that not all AI hype was created equal. Companies with high CRR (e.g. NVIDIA, Microsoft) largely justified their valuation surges through corresponding earnings and product breakthroughs, whereas those with low CRR (e.g. many startups and concept stocks) experienced the most pronounced booms and busts as expectations eventually met reality.

The case studies illustrate a continuum. On one end, pure-play AI firms like OpenAI became valuation juggernauts on promise alone, effectively serving as market anchors despite slender financials. On the other end, established firms like Adobe and even entire sectors like Chinese AI hardware saw bouts of exuberance corrected by the sobering force of fundamentals. Incumbents such as Meta and Microsoft demonstrated that strategic shifts toward AI, coupled with swift execution, can genuinely re-rate a business—and that doing so requires converting capability into tangible outcomes to avoid investor whiplash. The role of influential observers (Goldman Sachs) further reminds us that narratives at the macro scale can sustain or temper micro-level valuations.

A key takeaway is that \textbf{valuation misalignment risk} in emerging technologies can be gauged and, to some extent, managed. By monitoring indicators of CRR (for instance, AI-driven revenue as a percentage of total revenue, user engagement improvements due to AI, cost savings achieved via AI automation, etc.), investors and analysts can better distinguish between companies riding a fad versus those building sustainable competitive advantages. During the AI boom, many investors learned this the hard way—paying lofty multiples for “AI-themed” stocks that subsequently plunged. Our analysis suggests that a disciplined focus on realization (not just capability) could mitigate such risks.

Nonetheless, even in cases of overshooting, the long-term potential of AI to drive growth is real. The question is one of timing and distribution: the benefits will likely accrue unevenly and perhaps later than the peak of hype would suggest. Markets, driven by human psychology, will always grapple with finding the right anchor. What our study underscores is the importance of dynamically adjusting that anchor as new information arrives—be it a blockbuster earnings report or a cautionary lack of progress.

\bibliographystyle{plain}
\bibliography{sample-base}

\begin{thebibliography}{10}
\bibitem{FirstTrust}
Communication services sector performance since inception.
\bibitem{DLN}
Generative ai could automate almost half of all legal tasks - report.
\end{thebibliography}

\end{document}